\documentclass[12pt]{article}
\usepackage{cite}
\begin{document}
\title{Quark model relations for b-baryon decay}
\author{Jerrold Franklin\footnote{Internet address:
Jerry.F@TEMPLE.EDU}\\
Department of Physics\\
Temple University, Philadelphia, PA 19122
\date{}}
\maketitle
\begin{abstract}
Properties of b-baryon decay matrix elements (amplitudes) are derived in a nonsymmetric quark model without any use of SU(6), SU(3), or SU(2) (isotopic spin) groups.
Equalities  between pairs of matrix elements  are derived, and  $\Lambda-\Sigma$ mixing is used to calculate the branching ratio for the  transition $\Lambda_b\rightarrow \Sigma^0$.

\end{abstract}

\section{INTRODUCTION}

High energy accelerators are producing large numbers of heavy baryons whose decay processes 	are now being studied experimentally.  Recent papers studying heavy baryons have generally been based on the group SU(3)\cite{geng,dery}.  However, there is no good evidence for SU(3) symmetry in the quark model, and all of the experimentally satisfied predictions that use SU(3) can be derived in the quark model without introducing any internal group. 

Sum rules for baryon properties can be derived in the quark model by assuming the dominance  
of one and two-body interactions that are the same
 for each baryon in a particular sum rule.
This two-body, baryon independent, approach was introduced some time ago using SU(6) quark wave functions\cite{frt,rss}.
It was later shown to give the same results without introducing any form of SU(6), SU(3), or SU(2) symmetry  for the quark wave functions, in a `nonsymmetric quark model'\cite{jf1}.
Any predictions for baryons can be produced by using three-quark spin wave functions, 
 with no use of internal groups.

In section 2 of this paper, we describe the nonsymmetric quark model.  In section 3, we 
use the model to derive relations between matrix elements of b-baryons.
Section 4 is a review of $\Lambda-\Sigma$ mixing, which we then apply to the 
matrix element for a $\Lambda_b$ to $\Sigma^0$ transition, which
would vanish in the absence of the mixing.
Our results are summarized in Section 5.

\section{THE NONSYMMETRIC QUARK MODEL}

The nonsymmetric quark model is described in sections III-VI of reference\cite{jf1}. 
Here, we give a brief review of how it is used for the properties of spin one-half baryons.

A quark model wave function is given by
\begin{eqnarray}
q_1q_2q_3\chi_{_1}\quad{\rm or}\quad q_1q_2q_3\chi_{_0},
\end{eqnarray}
where $\chi_{_1}$ and $\chi_{_0}$ are the three-quark spin states
\begin{eqnarray}
\chi_{_1}&=& \frac{1}{\sqrt{6}}\left[2\uparrow\uparrow\downarrow-\uparrow\downarrow\uparrow
-\downarrow\uparrow\uparrow\right]
\label{s1}\\
{\rm and}\quad\chi_{_0}&=& \frac{1}{\sqrt{2}}\left[\uparrow\downarrow\uparrow
-\downarrow\uparrow\uparrow\right].
\label{s0}
\end{eqnarray}
The spin states correspond to a three-quark spin state of total spin $1/2$ with the first two quarks having spin 1 (for $\chi_{_1}$) or spin 0 (for $\chi_{_0}$).

The quark order is not arbitrary, but must be coordinated with the spin wave function. For the spin function $\chi_{_1}$ shown above,
any two identical quarks must be chosen as the first two quarks in the wave function. This is to implement the Pauli principle, requiring identical quarks to be symmetrical with respect to interchange.

The Pauli principle only applies to two identical quarks like $uu$ or $dd$, but not to 
different quarks like $ud$.
There is no `extended Pauli principle' because the $u$ and $d$ quarks are unrelated
in the nonsymmetric model.  They are {\it not} taken as two states of a single particle, as is presumed to implement the isotopic spin formalism.

 For `flavor-degenerate' baryons, composed of three different quark flavors, we choose the quark order to be in the order of their masses. This quark order 
 will minimize any mixing between flavor-degenerate baryons\cite{jf2}. 
 The quark order can be changed, but then the quark spin vectors 
must be changed along with the quark ordering to preserve the proper quark-spin coordination.  (This process will be seen in the examples below.)

No symmetrization of these wave functions is necessary or helpful. The Pauli principle and the quark-spin structure shown gives the same results as would any internal group symmetrization.

\section{b-BARYON DECAY MATRIX ELEMENTS}

In this section, we study decays of the form $B\rightarrow Y \;({\rm or}\; N)+S$, 
where B is a
  b-baryon decaying to a hyperon or a nucleon, plus a neutral object S with spin-parity $(1^-)$ (for instance, a photon, or the J$/\psi$ meson).  
The transitions are from a spin $\frac{1}{2}$ baryon to a spin $\frac{1}{2}$ baryon and a spin $1^-$
meson or photon. 
Parity is not conserved in the transitions $b\rightarrow d\;{\rm or}\;b\rightarrow s$, so there will be a
 scalar and a pseudoscalar ($\bf\sigma\cdot p$)
 decay amplitude for b-baryon decay.
 
The two types of transition matrix elements are
\begin{eqnarray}
\hspace*{-1in}{\rm scalar:}\hspace{.24in}
A(B\rightarrow Y+S)&=&\langle nsq,\eta |K_{bq}{\bf T_{bq}}|nsb,\xi\rangle,\\
\hspace*{-.6in}{\rm pseudoscalar:}\quad A(B\rightarrow Y+S)&=&\langle nsq,\eta |K'_{bq}{\bf T_{bq}}\sigma_z(3)|nsb,\xi\rangle,
\end{eqnarray}
where the quark  $q$ is either a $d$ or an $s$ quark, and $\eta$ and $\xi$ 
 are spin states.  $\bf T_{bq}$ is a transition operator that converts a $b$ quark
  to a $d$ quark or an $s$ quark.
The constants $K_{bq}$ and $K'_{bq}$ are assumed to be the same (separately) for each baryon 
connected by $\bf T_{bq}$.

 We will only be considering decays of the $\Xi_b$ and $\Lambda_b$ baryons, for which the first two quarks are in the spin zero state $\chi_0$.  Then the third quark, the $b$ quark that decays, will always be spin up.
This means that the spin operator, $\sigma_z(3)$, will just equal one, so the pseudoscalar matrix element will be the same as the scalar matrix element.
On the other hand, the $\Xi'_b$ has the first two quarks in a spin one state, so the $b$ quark can be either up or down, and 
$\sigma_z(3)$ would have to be considered for its decay.  

We first study decays where the quark flavor transition is $b\rightarrow d$. 
For the transition $\Xi_b\rightarrow \Sigma$, the decay matrix elements are
\begin{eqnarray}
 A(\Xi^0_b\rightarrow \Sigma^0+S)&=&\langle usd,\tilde\chi_{_1}|K_{bd}{\bf T_{bd}}|usb,\chi_{_0}\rangle\nonumber\\
&=&\frac{K_{bd}}{2\sqrt{3}}\langle\left[2\uparrow\downarrow\uparrow-\uparrow\uparrow\downarrow
-\downarrow\uparrow\uparrow\right]
\left[\uparrow\downarrow\uparrow
-\downarrow\uparrow\uparrow\right]\rangle\nonumber\\
&=&\frac{(2+1)K_{bd}}{2\sqrt{3}}=\frac{\sqrt{3}}{2}K_{bd}.
\label{A1}\\
\quad A(\Xi^-_b\rightarrow \Sigma^-+S)&=&\langle dsd,\tilde\chi_{_1}|K_{bd}{\bf T_{bd}}|dsb,\chi_{_0}\rangle\nonumber\\
&=&\frac{K_{bd}}{2\sqrt{3}}\langle\left[2\uparrow\downarrow\uparrow-\uparrow\uparrow\downarrow
-\downarrow\uparrow\uparrow\right]
\left[\uparrow\downarrow\uparrow
-\downarrow\uparrow\uparrow\right]\rangle\nonumber\\
&=&\frac{(2+1)K_{bd}}{2\sqrt{3}}=\frac{\sqrt{3}}{2}K_{bd}.
\label{A2}
\end{eqnarray}
Note that in these steps, the quark spin vectors for quark 2 and quark 3 had to be switched from spin state $\chi_{_1}$ to a transposed state $\tilde\chi_{_1}$
 to coordinate with the switch from $uds$ to $usd$ (and $dds$ to $dsd$) in the quark order.

 Comparing these two equations
leads to the relation
\begin{equation}
A(\Xi^0_b\rightarrow \Sigma^0+S)=A(\Xi^-_b\rightarrow \Sigma^-+S)
\end{equation}
for the decay matrix elements.
In DGGS, this relation is given as
\begin{equation}
A(\Xi^0_b\rightarrow \Sigma^0+S)=\frac{1}{\sqrt{2}}A(\Xi^-_b\rightarrow \Sigma^-+S),
\quad({\rm DGGS)}.
\label{is1}
\end{equation}
The difference in the two results is that DGGS uses the isotopic spin formalism where the $ud$ spin one quark state is given as $\frac{1}{\sqrt{2}}(ud+du)$, with $ud$ and $du$ considered as two different states.  This results in the factor $\frac{1}{\sqrt{2}}$ in Eq.~(\ref{is1}).

Although the DGGS result differs from ours for the matrix element\\\
 $A(\Xi^0_b\rightarrow \Sigma^0+S)$, it gives the same result for the decay rate.
The decay rate to the $\Sigma^0$ depends on $|A|^2$, for which the DGGS value is one half of ours.
But, in the isotopic spin formalism, the quark states $ud$ and $du$ are two different states,
each of which leads to a separate final decay state. This restores the isotopic spin decay rate to the same value as given by the nonsymmetric quark model.
 
Using the same procedure as above, the matrix elements for other decay modes of b-baryons 
for the transition $b\rightarrow d$ 
are  given by

\begin{eqnarray}
 A(\Xi^0_b\rightarrow \Lambda+S)&=&\langle usd,\tilde\chi_{_0}|K_{bd}{\bf T_{bd}}|usb,\chi_{_0}\rangle\nonumber\\
&=&\frac{K_{bd}}{2}\langle
\left[\uparrow\uparrow\downarrow-\downarrow\uparrow\uparrow\right]
\left[\uparrow\downarrow\uparrow
-\downarrow\uparrow\uparrow\right]
\rangle\nonumber\\
&=&\frac{(0+1)K_{bd}}{2}=\frac{K_{bd}}{2},
\label{A3}\\
A(\Lambda_b\rightarrow n+S)&=&\langle udd,\tilde\chi'_{_1}|K_{bd}{\bf T_{bd}}|udb,\chi_{_0}\rangle\nonumber\\
&=&\frac{K_{bd}}{2\sqrt{3}}\langle\left[2\downarrow\uparrow\uparrow
-\uparrow\downarrow\uparrow-\uparrow\uparrow\downarrow\right]
\left[\uparrow\downarrow\uparrow
-\downarrow\uparrow\uparrow\right]\rangle\nonumber\\
&=&\frac{(-2-1)K_{bd}}{2\sqrt{3}}=-\frac{\sqrt{3}}{2}K_{bd}.
\label{A4}
\end{eqnarray}

Combining Eqs.~(\ref{A1}), (\ref{A2}), (\ref{A3}), and (\ref{A4})  gives the relations
\begin{eqnarray}A(\Xi^0_b\rightarrow \Sigma^0+S)&=&A(\Xi^-_b\rightarrow \Sigma^-+S)\label{r1}\\
&=&\sqrt{3}A(\Xi^0_b\rightarrow \Lambda+S)\label{r2}\\
&=&-A(\Lambda_b\rightarrow n+S).\label{r3}
\end{eqnarray}

Our equality (\ref{r1}) agrees with the correponding equality (2.18) in DGGS 
when the isotopic spin factor $\frac{1}{\sqrt{2}}$ is taken into account.
The equalities  (\ref{r2}) and (\ref{r3}) differ only in sign with the corresponding equalities (2.20), (2.21), and (2.22) in DGGS.
This sign difference is probably due to different quark orderings or sign conventions,
and would not lead to a difference in branching ratios.

Decay matrix elements for the quark flavor transition $b\rightarrow s$ are 
\begin{eqnarray}
 A(\Xi^0_b\rightarrow \Xi^0+S)&=&\langle uss,\tilde\chi'_{_1}|K_{bs}{\bf T_{bs}}|usb,\chi_{_0}\rangle\nonumber\\
&=&\frac{K_{bs}}{2\sqrt{3}}\langle
\left[2\downarrow\uparrow\uparrow-\uparrow\downarrow\uparrow
-\uparrow\uparrow\downarrow-\right]
\left[\uparrow\downarrow\uparrow
-\downarrow\uparrow\uparrow\right]
\rangle\nonumber\\
&=&\frac{(-2-1)K_{bs}}{2\sqrt{3}}=-\frac{\sqrt{3}}{2}K_{bs},
\label{A5}\\
 A(\Xi^-_b\rightarrow \Xi^-+S)&=&\langle dss,\tilde\chi'_{_1}|K_{bs}{\bf T_{bs}}|dsb,\chi_{_0}\rangle\nonumber\\
&=&\frac{K_{bs}}{2\sqrt{3}}\langle
\left[2\uparrow\uparrow\downarrow-\uparrow\downarrow\uparrow
-\downarrow\uparrow\uparrow\right]
\left[\uparrow\downarrow\uparrow-\uparrow\uparrow\downarrow\right]
\rangle\nonumber\\
&=&\frac{(-2-1)K_{bs}}{2\sqrt{3}}=-\frac{\sqrt{3}}{2}K_{bs},
\label{A6}\\
 A(\Lambda_b\rightarrow \Lambda+S)&=&
\langle uds,\chi_{_0}|K_{bs}{\bf T_{bs}}|udb,\chi_{_0}\rangle\nonumber\\
&=&\frac{K_{bs}}{2}\langle
\left[\uparrow\downarrow\uparrow-\downarrow\uparrow\uparrow\right]
\left[\uparrow\downarrow\uparrow-\downarrow\uparrow\uparrow\right]
\rangle\nonumber\\
&=&(1+1)K_{bs}/2=K_{bs}.
\label{A7}
\end{eqnarray}
Combining Eqs.~(\ref{A5}), (\ref{A6}), and (\ref{A7}) gives the relations
\begin{eqnarray}
A(\Xi^0_b\rightarrow \Xi^0+S)&=&(\Xi^-_b\rightarrow \Xi^-+S)\label{r7}\\
&=&-\frac{\sqrt{3}}{2}(\Lambda_b\rightarrow \Lambda+S).
\label{r8}
\end{eqnarray}
Equation (\ref{r7}) agrees with Eq.~(2.15) in DGGS, and Eq.~(\ref{r8}) differs only in sign with their Eq.~(2.16) when the $\frac{1}{\sqrt{2}}$ isotopic spin factor is taken into account.

Because of the orthogonality of the $\Lambda_b$ and $\Sigma^0$ spin functions, the
$\Lambda_b\rightarrow \Sigma^0$
 transition matrix elements vanish:
\begin{eqnarray}
A(\Lambda_b\rightarrow \Sigma^0+S)&=&\langle uds,\chi_{_1}|K'_{bs}{\bf T_{bs}}|udb,\chi_{_0}\rangle\nonumber\\
&=&\frac{K_{bs}}{2\sqrt{3}}\langle
\left[2\uparrow\uparrow\downarrow-\uparrow\downarrow\uparrow-\downarrow\uparrow\uparrow\right]
\left[\uparrow\downarrow\uparrow-\downarrow\uparrow\uparrow\right]
\rangle\nonumber\\
&=&(-1+1)K_{bs}/2\sqrt{3}=0.
\label{r9p}\\
\end{eqnarray}
We show below that including $\Lambda-\Sigma$ mixing leads to a small non-vanishing branching ratio for the decay mode $\Lambda_b\rightarrow \Sigma^0+S.$

\section{LAMBDA-SIGMA MIXING}

There have been a number of calculations of $\Lambda-\Sigma$ mixing in the quark model\cite{jf2,ms,dvh,gs,dgg,isgurprd21,kordov}, 
most of which have relied on SU(3) symmetry and other assumptions.  
Here, we summarize the derivation of the $\Lambda-\Sigma$ mixing angle in \cite{jf2}, using
the nonsymmetric quark model.  The derivative is model independent in that it does not depend on any feature of the quark interactions, and makes no use of any group property.

The $\Lambda-\Sigma$ transition matrix element is
\begin{equation}
\epsilon=\langle uds,\chi_{_0}|H|uds,\chi_{_1}\rangle.
\end{equation}
Using the nonsymmetric quark wave functions, it is shown in \cite{jf2} that the transition matrix element
is given by
\begin{equation}
\epsilon=\frac{\sqrt{3}}{4}(D^0_{us}-D^1_{us}+D^1_{ds}-D^0_{ds}), 
\end{equation}
where $D^s_{ij}$ is a two-body interaction energy for the two labeled quarks, with the combined spin of the two quarks
being either 0 or 1.
The quark mixing angle $\theta_m$ is then given by
\begin{equation}
\tan(2\theta_m)=\frac{-2\epsilon}{(M_\Sigma-M_\Lambda)}.
\end{equation}

The combination ($D^0_{us}-D^1_{us}+D^1_{ds}-D^0_{ds})$
is also related to two sums of baryon masses, so $\epsilon$ can be given by either\cite{jf1}
\begin{eqnarray}
\epsilon_\Sigma&=&(M_{\Sigma^*_+}-M_{\Sigma^*_-}+M_{\Sigma_+}-M_{\Sigma_-})/2\sqrt{3}=-1.07\pm .02 {\rm MeV}
\end{eqnarray}
{\rm or}\footnote{The sum rule for $\epsilon_\chi$ was originally derived in reference \cite{gs} using SU(6) quark wave functions.}
\begin{eqnarray}
\quad\epsilon_\chi&=&(M_{\chi^*_+}-M_{\chi^*_-}+M_{\chi_+}-M_{\chi_-})/2\sqrt{3}=-1.07\pm 0.02 {\rm MeV}.
\end{eqnarray}
The agreement of these two measures of $\epsilon$ supports the assumption of baryon independence for the two-body interactions.
The baryon masses for these equations have been taken from the experimental summary of PDG\cite{PDG}.

Combining these two values for $\epsilon$, the $\Lambda-\Sigma$ mixing angle is given by
\begin{equation}
\theta_m=\frac{-\epsilon}{(M_\Sigma-M_\Lambda)}=\frac{1.07}{77}=0.014=0.80^\circ\pm 0.02^\circ.
\end{equation}
We have made a small angle approximation for $\theta_m$.
The value $\theta_m=0.80\pm 0.02^\circ$ is the mixing angle value that should be used for heavy baryon decays.

The mixing of the $\Lambda$ with the $\Sigma^0$ in $\Lambda_b\rightarrow \Sigma^0+S$ decay allows this decay mode to occur, even though their spin wave functions are orthogonal.
With $\Lambda-\Sigma$ mixing, the $\Lambda_b\rightarrow \Sigma^0$
matix element becomes
\begin{eqnarray}
A(\Lambda_b\rightarrow \Sigma^0+S)&=&0+\sin\theta_m\times A(\Lambda_b\rightarrow \Lambda+S).
\label{A7m}
\end{eqnarray}
Then the branching ratio for the rates of the two decays, including a phase space correction 
factor\cite{LHCb} $\Phi_{\Lambda_b}$, is
\begin{eqnarray}
{\cal R}&=&\Phi_{\Lambda_b}
\left|\frac{A(\Lambda_b\rightarrow \Sigma^0+S)}{A(\Lambda_b\rightarrow \Lambda+S)}\right|^2=\Phi_{\Lambda_b}\sin^2\theta\nonumber\\
&=&1.058(.014)^2=(2.1\pm 0.3)10^{-4}.
\label{A7n}
\end{eqnarray}

The LHCb collaboration has measured an upper limit of ${\cal R}<21\times 10^{-4}$ at $95\%$ confidence level\cite{LHCb}.
Our prediction is $10\%$ of that experimental upper limit.

\section{SUMMARY}

We have derived a number of relations between b-baryon decay amplitudes in a nonsymmetric quark model with no use of the SU(6), SU(3), or SU(2) groups.
All our predictions for branching ratios are well below current experimental limits, so tests of the predictions depend on improved measurement levels.

 We have used no group theory, but derived the same branching ratios as DGGS did by introducing the internal symmetry groups SU(3) and SU(2) (isotopic spin).   The group theory results are the same as ours because the group Clebsch-Gordan coefficients are the same as the spin addition coefficients seen in Eqs.~(\ref{s1})
and (\ref{s0}).  

Implementing SU(3) requires the three quark wave function to be completely symmetric in spin-space-SU(3) (and antisymmetric in color), which is only possible if the two sets of coefficients are the same.
This explains why SU(3) sum rules work well, even with quarks of differing masses.  It is not that SU(3) is a good symmetry, but that its 
Clebsch-Gordan coefficients must match those of the spin addition.


\begin{thebibliography}{9}
\bibitem{geng}C. Q. Geng, C.-W. Liu, and T.-H. Tsai, Phys.~Rev. D {\bf 101}, 054005 (2020), arXiv:2002.09583 [hep-ph].
\bibitem{dery}A. Dery, M. Ghosh,Y. Grossman, et al., J. High Energ. Phys. {\bf 165} (2020). https://doi.org/10.1007/JHEP03(2020)165\\
We refer to this paper as DGGS.
\bibitem{frt}P. Federmann, H. R. Rubinstein, and I. Talmi, Phys. Lett. {\bf 22}, 208 (1966). 
\bibitem{rss}H. R. Rubinstein, F. Scheck, and R. H. Socolow, Phys. Rev. {\bf 154}, 1608 (1967).
\bibitem{jf1}J. Franklin, Phys. Rev. {\bf 172}, 1807 (1968).\\
The `hidden spin' in the title of this paper was a precursor of `color',\\
 but the model in the paper is the usual quark model.
\bibitem{jf2}J. Franklin, D. B. Lichtenberg, W. Namgung, and D. Carydas,
Phys. Rev. D {\bf 24}, 2910 (1981).
\bibitem{ms}A. J. Macfarlane and E. C. G. Sudarshan, Nuovo Cimento {\bf 31}, 1176 (1964).
\bibitem{dvh}R. H. Dalitz and F. Von Hippel, Phys. Lett. {\bf 10}, 153 (1964).
\bibitem{gs}A. Gal and F. Scheck, Nucl. Phys. B2, 110 (1967).
\bibitem{dgg}A. De Rujula, H. Georgi, and S. L. Glashow, Phys. Rev. D {\bf 12}, 147 (1975).
\bibitem{isgurprd21}N. Isgur, Phys. Rev. D {\bf 21}, 779 (1980),
 [Erratum: Phys. Rev. D {\bf 23}, 817 (1981)].
\bibitem{kordov}Z. R. Kordov, R. Horsley, Y. Nakamura, H. Perlt, P. E. L. Rakow, G. Schierholz, H. Stüben,
R. D. Young, and J. M. Zanotti, (2019), arXiv:1911.02186 [hep-lat].
\bibitem{PDG}M. Tanabashi et al. (Particle Data Group), Phys. Rev. D{\bf 98}, 030001 (2018) and 2019 update.
\bibitem{LHCb}R. Aaij et al. (LHCb), Phys. Rev. Lett. {\bf 124}, 111802 (2020),
 arXiv:1912.02110 [hep-ex].
\end{thebibliography}
\end{document}